# Novel Use of Photovoltaics for Backup Spacecraft Laser Communication System

Xinchen Guo
The School of Electrical and
Computer Engineering
Purdue University
207 S. Martin Jischke Dr., IN 47906
xg@purdue.edu

Jekan Thangavelautham
Space and Terrestrial Robotic
Exploration (SpaceTREx) Lab
Arizona State University
781 E. Terrace Rd., AZ 85287
jekan@asu.edu

*Abstract*—Communication with a spacecraft is typically performed using Radio Frequency (RF). RF is a well-established and well-regulated technology that enables communication over long distances as proven by the Voyager 1 & II missions. However, RF requires licensing of very limited radio spectrum and this poses a challenge in the future, particularly with spectrum time-sharing. This is of a concern for emergency communication when it is of utmost urgency to contact the spacecraft and maintain contact, particularly when there is a major mission anomaly or loss of contact. For these applications, we propose a backup laser communication system where a laser is beamed towards a satellite and the onboard photovoltaics acts as a laser receiver. This approach enables a laser ground station to broadcast commands to the spacecraft in times of emergency. Adding an actuated reflector to the laser receiver on the spacecraft enables two-way communication between ground and the spacecraft, but without the laser being located on the spacecraft. In this paper, we analyze the feasibility of the concept in the laboratory and develop a benchtop experiment to verify the concept. We have also developed a preliminary design for a 6U CubeSat-based demonstrator to evaluate technology merits.

## TABLE OF CONTENTS

## 1. INTRODUCTION

Conventional spacecraft use radio frequency (RF) to communicate. RF technology enables long-distance communication and is reliable. The Deep Space Network (DSN) utilizes RF to both communicate and track dozens of space assets situated throughout the solar system. DSN offers high data-rates and high duty-cycle communication. However, RF spectrum is congested near earth due to ever increasing demand for high-bandwidth wireless communication. This has required strict licensing and regulation. With calls for spectrum time-sharing, such technology poses major concerns for space communication in the event of an emergency or during critical periods when a spacecraft is faced with technical anomalies.

Spacecraft communication can benefit from a backup communication system that requires very little onboard power and uses an alternate EM wavelength that eliminates interference while maximizing RF utilization. LADEE [1] spacecraft successfully demonstrated laser communication from Earth to the Moon (Figure 1). However, LADEE required diverting significant onboard power for communication and was operating at a low duty-cycle. This is a major concern for interplanetary missions where ever increasing spacecraft power needs to be diverted for communication with ground. A better solution is needed.

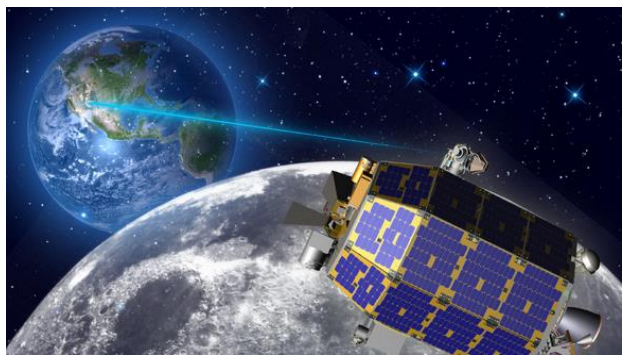

Figure 1: LADEE demonstrated the feasibility of high bandwidth laser communication from lunar orbit.

We propose a laser communication system where a laser is beamed towards a satellite and the onboard photovoltaics acts as a laser receiver. This approach enables a laser ground station to broadcast commands to the spacecraft in times of emergency. Adding an actuated reflector to the laser receiver on the spacecraft enables two-way communication between ground and the spacecraft, but without the laser diode being located on the spacecraft. Eliminating the laser onboard the spacecraft makes this technology applicable for low-power spacecraft, particularly CubeSats, PicoSats and FemtoSats. Advances in attitude control of CubeSats make it possible for them to achieve 10-20 arc second pointing accuracy, which is more than sufficient for laser communication requirements. In addition, this method avoids lifetime concerns of a laser diode onboard a spacecraft. However, the ground station will need to look behind or point-ahead to anticipate and maintain a link with a moving spacecraft. Furthermore, the ground station optics needs to be sensitive to detect a much weaker reflected beam. The technology is also conducive to close-range spacecraft to spacecraft communication.

Our proposed approach enables the laser communication system to be extensible. In other words, this system can enables future upgrades to the laser ground station, particularly (1) improvement in optics to both transmit and detect the reflected laser light, (2) increased power to the laser that effectively increases both bandwidth and range.

However we envision the first applications be as a backup communication system that can have very low bandwidth in the order 10 Kbps or less, but be sufficient to communicate telemetry and critical system states of spacecraft in Low Earth Orbit (LEO) and MEO (Medium Earth Orbit).

The bottleneck to data bandwidth comes from the laser signal filtering capabilities and reflector actuation response times of the spacecraft. Improvement in ground laser optics, increased power to the laser and larger reflector mirrors can address constraints on the signal filtering capabilities of the spacecraft. In addition, improvement to the optics and ground based detectors can be made. This enables the technology to increase its range, post spacecraft deployment. Several technologies are considered for mirror actuation on the spacecraft including DLPs, reflector LCDs, retro-modulators and thin-film quantum wells for high bandwidth applications [3, 5-6], while solar panel gimballing for low-bandwidth communication. Further improvements are possible from deploying relay satellites.

In this paper, we first review state-of-the-art in laser-communication in Section 2, followed by presentation of the proposed low-power, Extensible Laser Communication system (ELC) in Section 3. This will include a preliminary design, system specification, overview of the concept and theoretical feasibility studies. In Section 4, we present laboratory experiments performed to test some of the critical enabling technologies for ELC. This includes demonstration of a bench-top ELC proof-of-concept system, followed by discussions in Section 5, conclusions and future work in Section 6.

## 2. RELATED WORK

Miniaturization of space technology particularly electronics, CCDs and various science instruments have allowed for increased pixel scale resolution, giving higher fidelity data. This has translated into significant increase in the data generated and requires the communication data bandwidth to enable retrieval of this data to a ground station. The number of communication pipelines to a spacecraft beyond Low Earth Orbit is limited, often relying on a few ground stations such as the NASA Deep Space Network. These ground station have high costs associated with their maintenance and maintain priority for flagship missions. Alternate modes of communication that is not reliant on RF or the Deep Space Network open whole new opportunities for small spacecraft both through government and commercial entities.

Laser communication compared with traditional radio frequency communication methods can in theory provide much higher bandwidth with relatively small mass, volume and power requirements. This is possible because laser enable the beams of photons to be coherent over longer distances. LADEE demonstrated the advantages of laser communication, providing high bandwidth for a relatively small sized spacecraft [1]. However LADEE utilized laser system onboard the spacecraft to perform high-speed bidirectional communication and consumes between 50 and 120 Watts. This is too high for small spacecraft that typically produce a total power of less than 20 Watts.

In our approach, we eliminate the need for a laser on the spacecraft itself. This simplifies the bi-directional communication system on the spacecraft, significantly reducing its cost, mass, volume and power usage. The bandwidth maybe raised independent of the spacecraft by raising the transmit power of the ground station. This decouples the ground station from the spacecraft. However this requires modulating the incoming laser signal.

Generally, there are two kinds of technologies to modulate laser signal. One kind is mechanical steering mirror, which is produced by several companies. However, these technologies typically provide an update rate lower than 1 KHz [7]. The other kind, electro optical system could provide much higher update rates. A ferroelectric liquid crystal retromodulator was designed by Utah State University. The modulation rate was up to 20 Kbps; and a 1200 bps field test rate was achieved [2]. In order to achieve even higher speed, quantum well thin films were developed by Naval Research Lab to modulate laser beam [3, 5-6]. Early prototypes in the laboratory supported up to 600 kbps rate and 460 kbps rate was achieved in real experiment, which was limited by computer hardware at that time [3]. Another work shows that the quantum well device was developed to a rate of 1 Mbps [4]. In more recent paper, Naval Research Lab reported a field test result up to 10 Mbps [5], which demonstrated the device performance was of course better than the test result. These results show promise for laser modulation technologies onboard the spacecraft.

## 3. EXTENSIBLE LASER COMMUNICATION

Figure 2 shows the proposed system architecture. The ground station is equipped with a photo detector, a series of filters to process the light signal a microcontroller, a laser source and a series of direction actuators. Direction actuators are used to point the laser receiver and photo detector towards the target.

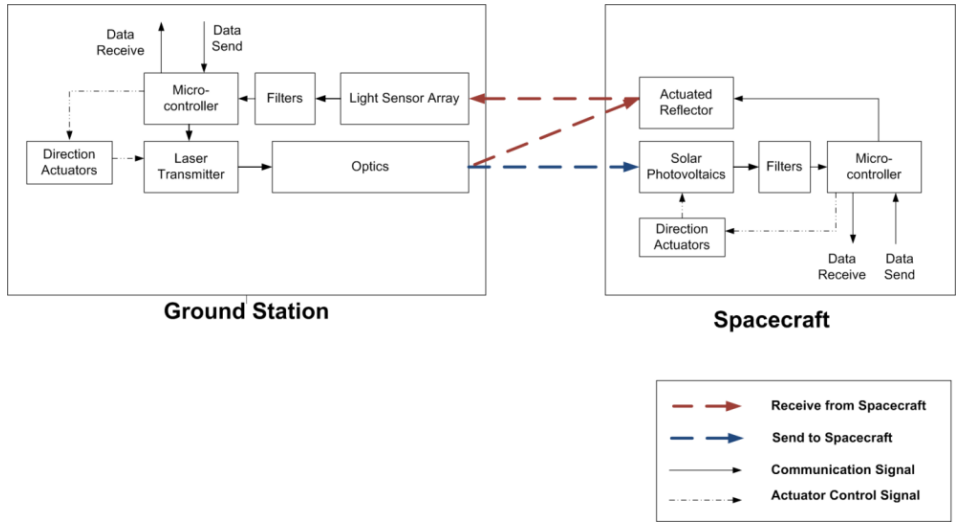

**Figure 2: Extensible Laser Communication System Architecture**

The received signal is filtered to gain maximum signal to noise ratio and the data acquired using a micro-controller which then transmit the binary data to other devices connected by wire to the ground station. The laser source is only located on the ground station and can send modulated laser beam to the spacecraft or robot.

On board the spacecraft or robot is the laser beam detector, which would nominally be a large solar photovoltaic panel or even photodetector, a bank of filters to process the signal from the photovoltaics, an actuated reflector, microcontroller and some attitude or direction actuator to point the reflector and detector. Data is sent from the ground station simply with the laser beam being beamed to the spacecraft or robot. For the spacecraft or robot to send data to the ground station, the ground station send a continuous laser beam to the spacecraft or robot that would then be modulated and reflected back by a high-speed switching reflector onboard the platform. This reflector can be a simple mechanical actuated mirror, Texas Instrument's DLP, reflective LCD or quantum well technology [3, 5-6].

A simple communication protocol to receive data from the spacecraft is shown in Figure 3. The spacecraft intends to send data shown on the first row (spacecraft send). The laser on the ground is modulated to "high" and beamed to the spacecraft. The spacecraft reflects back the laser signal (labeled ground receive). Sending data to the spacecraft is very straightforward with the laser modulated to beam the binary signals from the ground to the spacecraft (Figure 4).

In the following section we analyze the performance of laser communication technologies by using the Apache Point Observatory (APO) that perform lunar ranging as reference. Plugging this method into existing ranging and optical communication projects, we could get the minimum requirement for laser ranging and optical communication respectively.

In this method, the signal strength is believed to be equal to the product of emitted signal strength and ratio of receiving mirror area and area of laser spot projected on the moon. Since area of laser spot increases quadratically with the working distance, emitted signal strength is thought to be fixed, which only has linear effect.

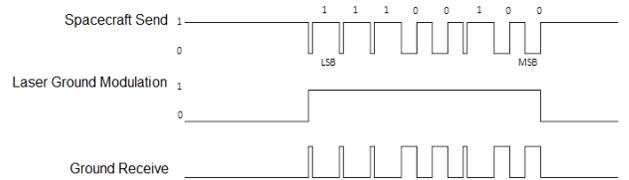

**Figure 3: Receive data from spacecraft.**

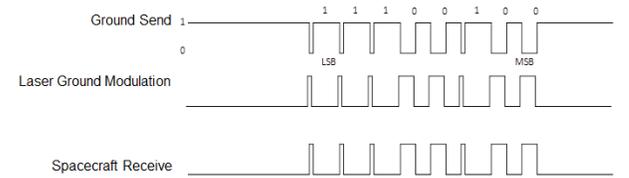

**Figure 4: Send data to ground.**

According to the estimation method mentioned above, we could get the expression of area ratio, which has arbitrary unit. We take LLCD project as a reference to estimate the baseline of optical communication. The receiver consists of 4 telescopes on earth and the space segment consists of 1 telescope to emit laser beam.

$$R_{com} = \frac{4d_{tele}^2}{\left(d_{me}\theta_{div}\right)^2} \qquad (1)$$

Where $R_{com}$ is the area ratio; $d_{tele}$ is the diameter of receiving telescope; constant 4 denotes that the receiver is an array of 4 telescopes; $d_{me}$ is the distance between moon and earth and $\theta_{div}$ is the angle of laser divergence. First, we set the distance between moon and earth $d_{me}$ as 384,400 km. The diameter of receiving telescope $d_{tele}$ is 0.40m. The divergence of laser beam $\theta_{div}$ is assumed to be 1 arc-second, i.e. 4.848 µrad. Plugging above data into Equation 1, we get the ratio $R_{com}$ to be 1.8E-7. Similarly, we could yield area ratio expression for laser ranging. We take lunar ranging with APO telescope and the reflecting mirror left by Apollo 11 as references. The transceiver consists of 1 telescope on earth and the reflecting mirror consists of 100 small mirrors.

$$R_{ran} = \frac{100 d_m^2 d_{tele}^2}{\left(d_{me}\theta_{em}\right)^2 \left(d_{me}\theta_{me}\right)^2} \quad (2)$$

Where $R_{ran}$ is the area ratio; $d_m$ is the diameter of one small mirror 3.8 cm; $d_{me}$ is the distance between moon and earth 384,400 km; $\theta_{em}$ is the divergence of laser beam from earth to moon 1 arc-second; $\theta_{me}$ is the divergence of laser beam from moon to earth 8 arc-second and $d_{tele}$ is the diameter of ground transceiver telescope 3.5 m. Thus we get $R_{ran}$ to be 2.29E-15. Based on these above calculations, we develop a minimum criterion for optical communication.

$$\frac{A_{mi} A_{tele}}{\left(\frac{\pi d \theta_1}{2}\right)^2 \left(\frac{\pi d \theta_2}{2}\right)^2} \geq R_{com} \quad (3)$$

Where $A_{ml}$ is the reflecting mirror area; $A_{tele}$ is the area of telescope on the ground; $d$ is the desired communication distance; $\theta_1$ is divergence of laser beam from earth, typically 1 arc-second; $\theta_2$ is the divergence of laser beam back to earth, typically 3 arc-second. We set the distance to be 400km on LEO. The required area of the reflecting mirror is shown in Figure 5.

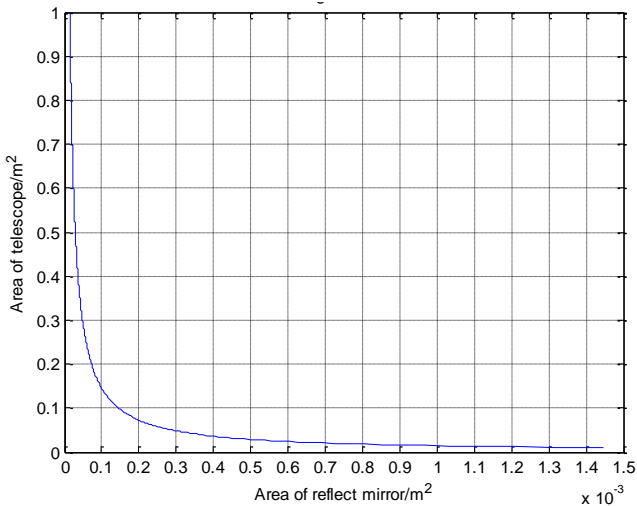

**Figure 5: Size of telescope required to detect signal from a reflector mirror in LEO.**

A telescope with a diameter of 0.4 m has the area of 0.126 m$^2$, which corresponds to a reflect mirror with 1.15E-4 m$^2$. This area is about 2 reflecting mirrors, which is about 5mm in diameter and developed by NRL. For the moon, we set the distance to be 384,400 km. Then the required area of reflecting mirror is calculated with telescope area from 1 m$^2$ to 10 m$^2$ and is shown in Figure 6.

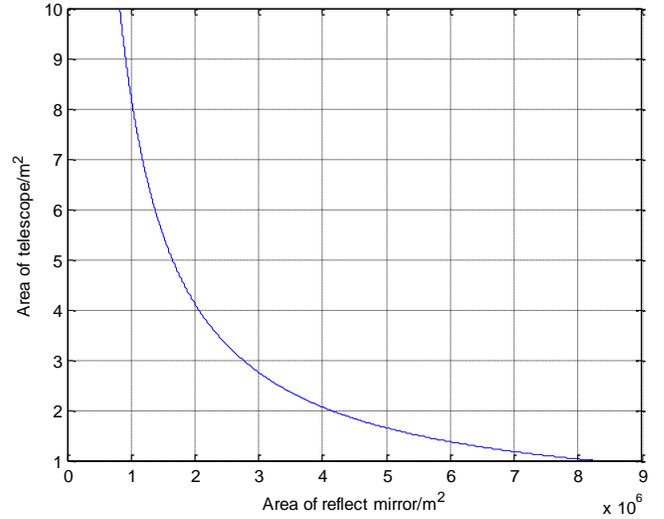

**Figure 6: Size of telescope required to detect signal from a reflector mirror in lunar orbit.**

From Figure 6, we could find a very different situation. If we use a telescope of 3.5m (about 10m$^2$) in diameter on earth, we still need a reflecting mirror about 2.5E6 m$^2$. We set the reflecting mirror to be 0.1 m, 0.2 m, 0.3 m and 0.4 m in diameter. Then the maximum distance is calculated with respect to different telescope area, ranging from 1 m$^2$ to 10 m$^2$ and is shown in Figure 7.

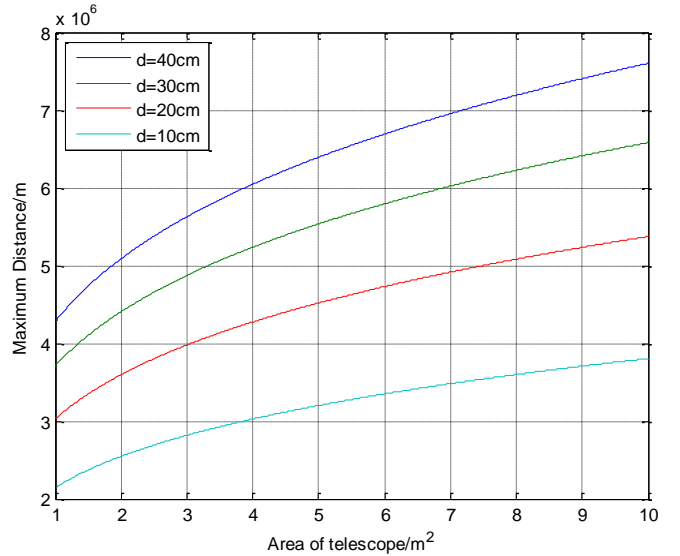

**Figure 7: Maximum communication distance achieved for telescope and reflector mirror combination.**

We could find that the realistic working distance is on the order of 10E6 m. That is to say, our proposed system is suitable for all LEO and lower MEO applications. To further analyze the feasibility of the proposed approach, we calculated maximum distance with the reflecting mirror up to 40 times the area of the Apollo 11 mirror located on the lunar surface shown in Figure 8. From Figure 8 we could find that even with an 80 m telescope and a 40x reflecting mirror, we could only achieve 2.7E9 m which is less than distance from Earth to the Moon.

Use of on-orbit laser relay satellites has not been explored in this paper, but this maybe sufficient for Earth to Moon communication. The challenge comes from having to install large detection mirrors or providing high enough power so that the reflected beam is detectable.

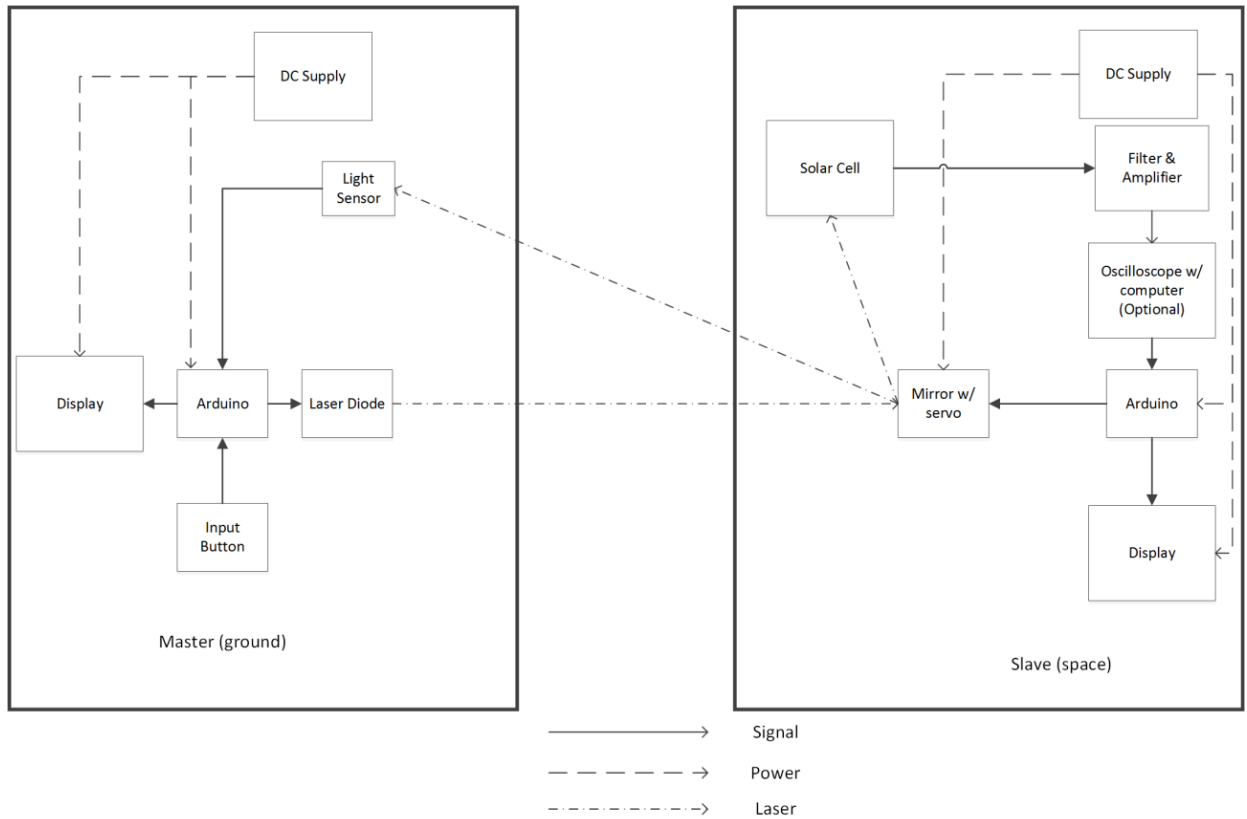

**Figure 9: Benchtop experiment setup to demonstrate the Extensible Laser Communication (ELC) system.**

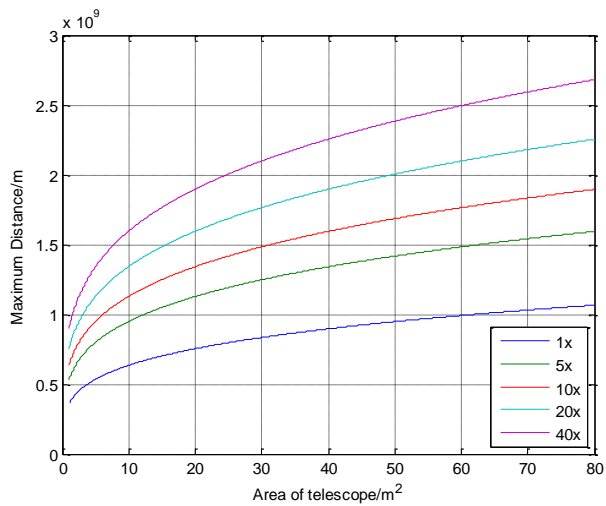

**Figure 8: Max comms. distance for telescope/reflector.**

## 4. EXPERIMENTS

Based on this proposed system, we developed a laboratory proof of concept demonstration of this proposed system (Figure 9, 10). On the spacecraft is the laser beam receiver that consists of a photovoltaic panel, a bank of filters and amplifiers to detect the laser signal and microcontroller. In this configuration, the ground station terminal and spacecraft terminal are separated by less than 0.2 m. In a real world system, the laser beam would consume much higher power and be focused using a series of optic to ensure coherence over hundreds of thousands of kilometers. The ability to send and receive laser signal over long distances (as far as the moon) have already been demonstrated from ranging experiments done since 1962.

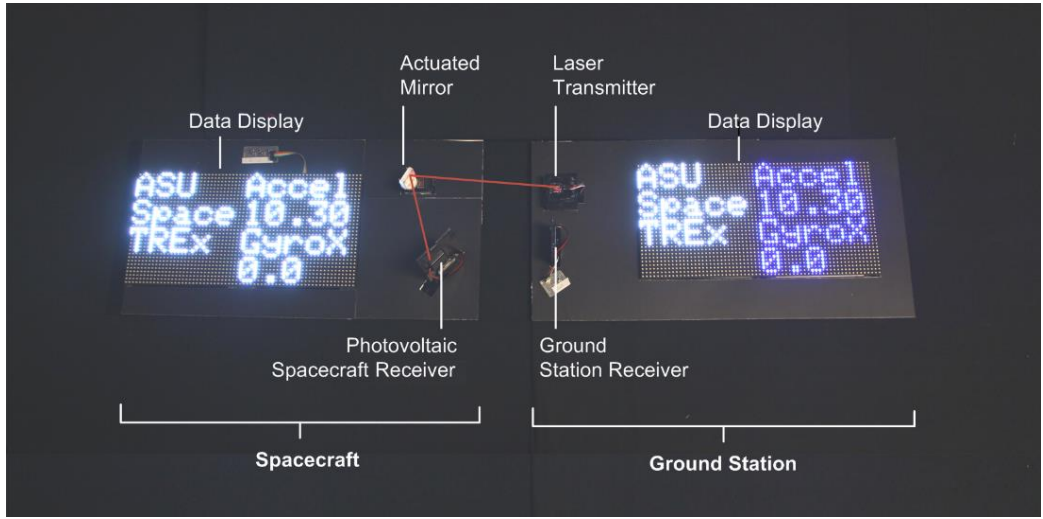

**Figure 10: Photo of benchtop system in operation.**

The proposed system was tested outside of the laboratory to evaluate whether the proposed laser communication approach, i.e. dual use of solar panels for producing solar power and for being large receivers for laser communication is possible. This was done by utilizing a violet laser. In July, Phoenix day, in the shade, a square wave signal from the violet laser is shown in Figure 11. Based on these signals detected (peak to peak of 58 mV) it is clear that with the right digital signal processing equipment it is possible to read the signal and perform laser communication. Figure 12 shows the same experiment repeated in sunlight (July, Phoenix).

The square signal from the laser is once detectable but amplitude has decreased to peak to peak of 11 mV, still more than sufficient for the right digital equipment to discern the signal from background sunlight.

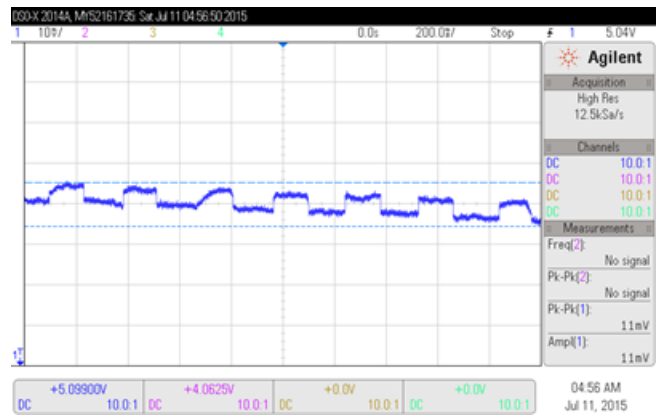

**Figure 12: Purple laser signal detected using Emcore triple junction cell outdoors in Phoenix, Az (July).**

Our laboratory experiments show the proposed laser communication system able to successfully communicate data between the simulated ground station and spacecraft terminals. The word "ASU SpaceTREx," together with data from accelerometer and gyro were successfully transmitted and received. A data rate of 10 KBps was achieved. The data rate was limited by the actuated mirror which composed of a servo attached to a mirror. This simulates the condition on a satellite, where the solar panels are independently gimballed. In a real world system, we would use one of the several proposed actuated reflectors to achieve high data rates if required. Further, we have tested the laser system on the solar photovoltaics under outdoor conditions. Under July sunlight in phoenix, the effective amplitude of the signal decreases, however the signal was detected. Overall these experimental demonstrate the principal feasibility of the technology for short distances in the laboratory.

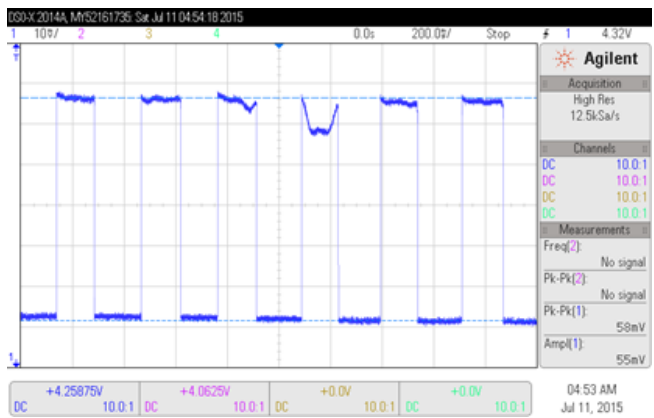

**Figure 11: Purple laser signal detected using Emcore triple junction cell in the shade in Phoenix, Az (July).**

## 6. DISCUSSION

Laser communication holds great potential as this technology can overcome bandwidth congestion and licenses issues faced with RF communication. Our study shows the potential opportunities for this technology for small spacecraft where there is limited power available. The major advantage of this technology is that the laser is located on the ground station. The laser maybe upgraded in the future and its beam coherence improved and that translates into improved communication performance and communication range for the spacecraft. The technology is ideally suited for small, low-power spacecraft in LEO and MEO. One proposed application could be to communicate and coordinate with satellites that monitor threats to Earth including incoming meteors and solar particles [8]. Another application is to use it on an on-orbit centrifuge science laboratory [9-11]. Table 1 show a comparison with conventional high data-throughput UHF technology. The technology however is applicable as backup communication system for spacecraft located in MEO and even GEO.

**Table 1: ELC Applied in LEO**

| Low Earth Orbit | Proposed System | UHF |
|---|---|---|
| Distance | 400 km | 400 km |
| Max Data Rates | 1 MBps | 3 MBps |
| Power Usage | 0.2 W | 25 W |
| Mass on Spacecraft | 0.2 kg | 0.25 kg |
| Cost on Spacecraft | $7,000 | $25,000 |

The proposed technologies utilize existing components on a spacecraft for the purpose of communication. This includes using solar PV as a laser receiver. Solar panels happen to have the largest area of any component or device on a spacecraft and can be accurately pointed. Thus a spacecraft has the capability to maximize optical reception using existing subsystems including attitude control and solar panel gimballing capability. In addition the ability to use retroreflectors to reflect laser light over long distances is a proven technology. This technology has been proven to work through laser ranging experiments performed with the mirrors located on the Moon.

The potential is there for high speed communication at low power, thanks to selection of a fast, low-power actuated reflector. Interplanetary applications face some important hurdles with this technology as the telescope mirror and reflector mirror need to be large and this poses important practical challenges. One possibility is to have low-cost relay satellites, analogous to cell phone towers that would collect the incoming laser signal and boost the signal, while also increasing its coherence. The advantage of using laser light instead of RF is that high data rates may be achieved for low power of the end terminal. All of this, however, requires an in-space communications infrastructure.

Our work is now progressing towards developing a CubeSat demonstrator to evaluate this concept from LEO. Implementing the proposed technology on a CubeSat is a low-cost approach to both validate the technology in the field and evaluate some of the practical challenges of in-space deployment and operation. The proposed spacecraft is shown in Figure 13. It will compose of solar panels that will be able to detect laser light from a ground station. The panels will be gimballed to reflect back the laser light to ground. Additional options are being evaluated including use of an electro-reflective layer over the spacecraft solar panel. All of these results presented here point towards a promising pathway forward for further technology maturation and demonstration from space.

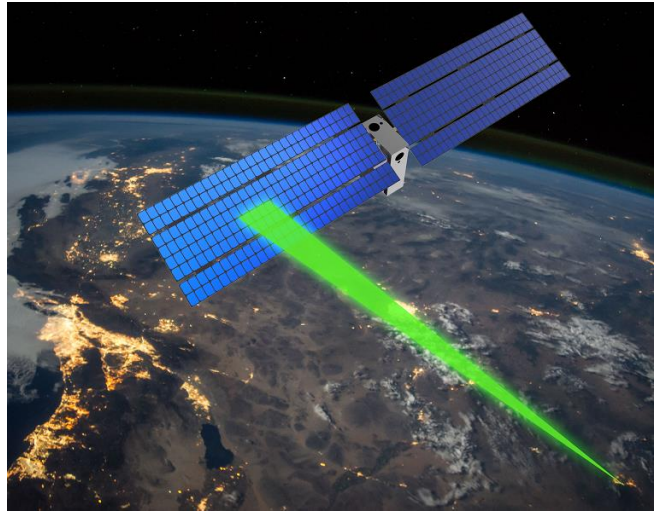

**Figure 13: 6U CubeSat concept to demonstrate the Extensible Laser Communication System from LEO.**

## 7. CONCLUSIONS

In this paper we have proposed a laser communication system for spacecraft that utilizes solar panels as receivers. The receiver in turn reflects the laser beam back towards ground to enable two-way communication. The technology enables low power communication and is best suited as a backup communication system in place of RF. However, with future advancement, this technology could be a primary form of communication that enables high-bandwidth communication for low power consumption aboard a spacecraft. The advantage of this technology is that the laser needs to be located at the ground station. This frees the spacecraft from having to host the laser system and high-power required. Furthermore, this technology is extensible as future upgrades to the laser and the optics on the ground will increase both performance and range of the spacecraft communication system. Our analysis shows this technology is feasible for communicating with spacecraft in LEO and MEO orbits. However, significant advancements are required to make this practical for interplanetary communication, as the size of the detecting telescope and reflector mirror increases significantly. Use of relay satellites may help to alleviate this challenge.

## ACKNOWLEDGEMENTS

The authors would like to thank Dr. Phil Dowd of AzTE for the helpful discussions and suggestions. In addition, the authors would like to thank the reviewers for their comments and suggestions. Artistic renderings of the concept made by Gavin Liu of ASU SpaceTREx.

## BIOGRAPHY

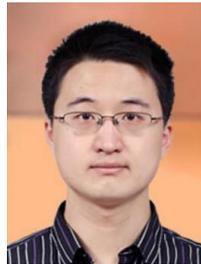

*Xinchen Guo received a B.E. in Materials Science and Engineering from University of Science and Technology Beijing, China in 2013. He received a M.S. in Materials Science and Engineering from Arizona State University in 2015. Now he is pursuing a PhD in Electrical and Computer Engineering in Purdue University. He developed a laser communication system in SpaceTREx laboratory during his study in ASU. Before joining ASU, he spent four years as a core member and then team lead in Chinese autonomous driving module car racing competitions.*

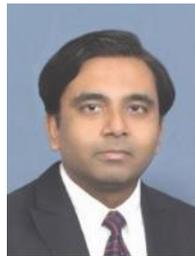

*Jekan Thangavelautham is an Assistant Professor and has a background in aerospace engineering from the University of Toronto. He worked on Canadarm, Canadarm 2 and the DARPA Orbital Express missions at MDA Space Missions. Jekan obtained his Ph.D. in space robotics at the University of Toronto Institute for Aerospace Studies (UTIAS) and did his postdoctoral training at MIT's Field and Space Robotics Laboratory (FSRL). Jekan Thanga heads the Space and Terrestrial Robotic Exploration (SpaceTREx) Laboratory at Arizona State University. He is the Engineering Principal Investigator on the AOSAT I CubeSat Centrifuge mission and is a Co-Investigator on SWIMSat, an Airforce CubeSat mission to monitor space threats.*